\documentclass[12pt,letter]{article}
\usepackage{epsfig,amsfonts,amssymb,graphicx}
\newcommand{\beq}{\begin{equation}}
\newcommand{\eeq}{\end{equation}}
\newcommand{\be}{\begin{equation}}
\newcommand{\ee}{\end{equation}}
\newcommand{\beqa}{\begin{eqnarray}}
\newcommand{\eeqa}{\end{eqnarray}}
\newcommand{\beqar}{\begin{eqnarray*}}
\newcommand{\eeqar}{\end{eqnarray*}}

\begin{document}

\begin{titlepage}

\begin{flushright}
DCPT-09/15\\
URB/35686
\end{flushright}

\vspace{3cm}

\begin{center}
{\bf \Large Geometry of the doubly spinning Black Ring}

\vspace{1cm}

Matthew Carlton

\vspace{0.5cm}

{\small\textit{Department of Mathematical Sciences, University of Durham}}\\
{\small\textit{Science Laboratories, South Road, Durham, DH1 3LE, United Kingdom}}

\vspace{0.5cm}

{\tt m.s.carlton@durham.ac.uk}
\end{center}

\vspace{1cm}

\begin{abstract}

We provide a detailed analysis of the doubly spinning black ring, investigating both its general properties and its shape.  We also examine the geometry of the ergosurface, illustrating the process of self-merging and discussing the physics of ergoregions.

\end{abstract}

\end{titlepage}

\newpage

\tableofcontents

\setcounter{footnote}{0}
\setcounter{equation}{0}
\section{Introduction}
\label{sec:intro}

In recent years, it has become apparent that black holes, the fundamental objects of General Relativity, admit a variety of exotic configurations in higher dimensions.  The discovery of the singly spinning black ring \cite{letter} by Emparan and Reall illustrated that black hole uniqueness does not hold in five spacetime dimensions: a black hole cannot be uniquely determined solely by its mass and angular momenta.

Since the initial discovery of the singly spinning black ring, the overall picture in five dimensions has become much clearer, largely due to a better understanding of the solution generating techniques involved and in particular the inverse scattering method.  As a result, black rings have since been constructed with solely $S^2$ angular momentum \cite{twosphere}, followed by the most generalised ring with two angular momenta \cite{PS}.  From this point, systems have been constructed with multiple black objects, including bi-rings \cite{bicycles}, di-rings \cite{diring}, and black saturns \cite{saturn}.

This progress has resulted in a more complete picture of the phase diagram for five dimensional black holes, yet many geometrical features of these solutions such as the shape of the horizon and ergosurface have not been so fully explored.  Many of the difficulties here lie with the problems inherent in visualizing a curved spacetime: nevertheless, Elvang, Emparan and Virmani provided a significant step in this direction through the technique of isometric embedding, in the process illustrating that the $S^2$ of black rings with $S^1$ angular momentum can admit severe distortion \cite{dynamics}.

In this paper we proceed similarly using the Pomeransky and Sen'kov balanced doubly spinning ring metric, finding the shape of the event horizon and calculating its relative location in space.  Another key geometrical feature of any rotating black hole is its ergosurface; the boundary of the region of spacetime which disallows static observers.  A recent paper \cite{ergoregion} discovered a remarkable topological transition between `unmerged' $S^1 \times S^2$ and `merged' $S^3$ ergosurfaces for the doubly spinning ring, depending on a simple condition of the parameters.  We illustrate this process here in more intuitive coordinates, and then go on to examine the shape of the ergosurface in the singly spinning case and discuss the associated physics.

Briefly, the structure and results of the paper are as follows:

\begin{itemize}

\item {\bf Parameters of the solution (section \ref{sec:parameters})}:
\begin{itemize}
When the mass of black rings is fixed, it is often most intuitive to think of the various configurations in terms of angular momenta, or equivalently, their horizon shape.  The metric and corresponding physical quantities are expressed in terms of three parameters, $k$, $\nu$ and $\lambda$, where $k$ sets a scale for the solution and $\nu$ and $\lambda$ can be thought of as parametrizing the angular momenta.  As such, we illustrate the population of the phase diagram in terms of these parameters so that the various black rings can be thought of readily in terms of $\nu$ and $\lambda$.
\end{itemize}

\item {\bf Geometry of the ring (section \ref{sec:geometry})}:
\begin{itemize}
$S^2$ angular momentum causes further distortion on the horizon two-sphere.  Very thin rings have circular cross sections only in the singly spinning ring limit: the $S^2$ rotation causes the horizon to become oblate, similarly to the horizon of a Kerr black hole.   We also find that very thin\footnote{We mention here that we also associate the description `thin' to the $S^1$ radii of the ring horizon.  Unlike in the singly spinning case, black rings exist whose inner and outer radii are arbitrarily close to each other yet have oblate two-spheres, and thus potentially large horizon areas.} black rings with large $j_\phi$ angular momenta exist much closer to the $S^1$ axis of rotation.
\end{itemize}

\item {\bf Physics of black ring ergosurfaces (section \ref{sec:physics})}:
\begin{itemize}
In the first subsection, we provide a simple derivation of the merger condition and illustrate it in the polar coordinates from which ring coordinates are constructed.  Physically, this is once again the result of the Kerr-like influence of the two-sphere, where the ergosurface bulges away from the horizon.  We then examine the shape of the ergosurface in the case of the singly spinning ring, and associate the relative shape and size of the ergosurface to the linear velocity of the horizon in the plane of the ring.
\end{itemize}

\end{itemize}

\section{Parameters of the solution}
\label{sec:parameters}

The metric is \cite{PS}\footnote{We adopt a metric of mostly positive signature, and take ring coordinates as in \cite{ringcoords}. Hence $\psi$ and $\phi$ are exchanged relative to \cite{PS}.}

\beqa
	ds^2 &=& -\frac{H(y,x)}{H(x,y)}(dt+\Omega)^2 
	-\frac{F(x,y)}{H(y,x)}d\psi^2 
	-2\frac{J(x,y)}{H(y,x)}d\psi d\phi 
	+\frac{F(y,x)}{H(y,x)}d\phi^2 \nonumber\\
	&& +\frac{2k^2H(x,y)}{(x-y)^2(1-\nu)^2}\left(\frac{dx^2}{G(x)}-\frac{dy^2}{G(y)}\right)
\eeqa          
where
\beqa
	\Omega &=& -\frac{2k\lambda\sqrt{(1+\nu)^2-\lambda^2}}{H(y,x)}[(1-x^2)y\sqrt{\nu}d\phi \nonumber\\
	&& +\frac{1+y}{1-\lambda+\nu}(1+\lambda-\nu+x^2y\nu(1-\lambda-\nu)+2\nu x(1-y)) d\psi]
\eeqa
and the functions G, H, J and F are specified by
\beqa
	G(x) &=& (1-x^2)(1+\lambda x+\nu x^2), \nonumber\\
	H(x,y) &=& 1+\lambda^2-\nu^2+2\lambda\nu(1-x^2)y+2x\lambda(1-y^2\nu^2)+x^2y^2\nu(1-\lambda^2-	\nu^2), \nonumber\\
	J(x,y) &=& \frac{2k^2(1-x^2)(1-y^2)\lambda\sqrt{\nu}}{(x-y)(1-\nu)^2}(1+\lambda^2-\nu^2+2(x+y)\lambda	\nu-xy\nu(1-\lambda^2-\nu^2)), \\
	F(x,y) &=& \frac{2k^2}{(x-y)^2(1-\nu)^2}[G(x)(1-y^2)[((1-\nu)^2-\lambda^2)(1+\nu)\nonumber\\
	&& +y\lambda(1-\lambda^2+2\nu-3\nu^2)]+G(y)[2\lambda^2+x\lambda((1-\nu)^2+\lambda^2)\nonumber	\\
	&& +x^2((1-\nu)^2-\lambda^2)(1+\nu)+x^3\lambda(1-\lambda^2-3\nu^2+2\nu^3)\nonumber\\
	&& -x^4(1-\nu)\nu(-1+\lambda^2+\nu^2)]]. \nonumber
\eeqa
The parameters $\nu$ and $\lambda$ satisfy the following constraints:
\be
	0 \le \nu < 1, \qquad
	2\sqrt{\nu} \le \lambda < 1+\nu.
\ee
The lower bound on $\lambda$ corresponds to extremal rings: that is, zero temperature black rings with maximal $j_\psi$ for given $j_\phi$.  In the limit $\lambda\rightarrow1+\nu$, we obtain the extremal Myers-Perry black hole \cite{bicycles}.\\
\\
The coordinate ranges are
\be
	-1 \le x \le 1, \qquad
	-\infty < y \le -1.
\ee
By taking $\nu \rightarrow 0$, setting $R^2=2k^2(1+\lambda^2)$ and then replacing $\lambda$ by $\nu$, we recover the black ring metric with angular momentum only on the $S^1$ \cite{ringcoords}.\\
\\
The zeros of $G(y)$, 
\be
	1+\lambda y+\nu y^2=0,
\ee
determine the locations of the inner and outer horizons.  Hence we see that an event horizon of topology $S^1 \times S^2$ lies at
\be
	y_h=\frac{-\lambda+\sqrt{\lambda^2-4\nu}}{2\nu}.
\ee
The mass and angular momentum are \cite{PS,bicycles}
\beqa
	M&=&\frac{3\pi k^2}{G_N}\frac{\lambda}{1+\nu-\lambda},\\
	J_\psi&=&J_1=\frac{2\pi k^3}{G_N}\frac{\lambda(1+\lambda-6\nu+\nu\lambda+\nu^2)\sqrt{(1+\nu)^2-	\lambda^2}}{(1+\nu-\lambda)^2(1-\nu)^2},\\
	J_\phi&=&J_2=\frac{4\pi k^3}{G_N}\frac{\lambda\sqrt{\nu[(1+\nu)^2-\lambda^2]}}{(1+\nu-\lambda)(1-\nu)^2}.
\eeqa
The angular velocities are given by
\beqa
	\Omega_\psi&=&\frac{1}{2k}\sqrt{\frac{1+\nu-\lambda}{1+\nu+\lambda}},\\
	\Omega_\phi&=&\frac{\lambda(1+\nu)-(1-\nu)\sqrt{\lambda^2-4\nu}}{4k\lambda\sqrt{\nu}}\sqrt{\frac{1+	\nu-	\lambda}{1+\nu+\lambda}},
\eeqa
and the horizon area and temperature are
\beqa
	\mathcal{A}_H&=&\frac{32\pi^2k^3\lambda(1+\nu+\lambda)}{(1-\nu)^2(y_h^{-1}-y_h)},\\
	T_H&=&\frac{(y_h^{-1}-y_h)(1-\nu)\sqrt{\lambda^2-4\nu}}{8\pi k\lambda(1+\nu+\lambda)}.
\eeqa
We fix a value for $k$ in terms of $\nu$ and $\lambda$ by setting $G_NM=1$.  We also introduce dimensionless quantities for the angular momentum and horizon area,
\be
	j_i=\sqrt{\frac{27\pi}{32G_N}}\frac{J_i}{M^{3/2}}, \qquad
	a_H=\frac{3}{16}\sqrt{\frac{3}{\pi}}\frac{\mathcal{A}_H}{(G_NM)^{3/2}},
\ee
and find that the `reduced' angular momenta are subject to the following conditions,
\be
	j_\phi\le\frac{1}{4}, \qquad
	j_\psi\ge\frac{3}{4}.
\ee
\\
Having established the physical quantities for the solution, we now wish to develop a better understanding of its parameters.  When working with black rings it is often easiest to classify them by their angular momenta, and consequently, their horizon shape.  As such, we illustrate the relation between $\nu, \lambda$ and $j_\psi, j_\phi$, outline the various classes of black ring of the resulting phase diagram, and provide an interpretation of the parameters.

\begin{figure}[h]
\begin{center}
\includegraphics[width=15cm]{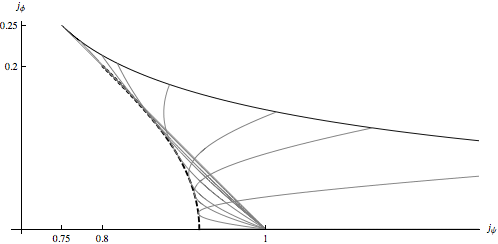}
\end{center}
\caption{\small Illustrative population of phase space of doubly spinning black rings for $j_\psi, j_\phi\ge0$.  The grey curves are branches of constant $\nu$, their values (from bottom to top) being 0.001, 0.005, 0.01, 0.03, 0.07, 0.1, 0.2, 0.95.  The solid black curve represents extremal zero temperature black rings.  The dashed black curve is the curve of cusps: doubly spinning rings with maximal horizon area and minimal $j_\psi$ for given $j_\phi$.}
\label{figure:phase1}
\end{figure}

Given the dependency of the bounds of $\lambda$ on $\nu$ it is logical to proceed by plotting curves of constant $\nu$ in $j_\psi, j_\phi$ space, allowing $\lambda$ to vary over the permitted constraint range.  A sample of the resulting curves are given in figure \ref{figure:phase1}.  Two pertinent features of the phase diagram are:

\begin{itemize}

\item
The vast majority of the doubly spinning ring phase diagram is given by a comparatively small range of $\nu$.  This is a result of the constraint on $\lambda$: as $\nu$ increases, the range of $\lambda$ quickly decreases.

\item
For every fixed value of $\nu$ we find both thin and fat rings.  The thin ring branch is given by the part of the curve lying between the curve of cusps (dashed curve) and the curve of extremal rings (black curve), while fat rings occupy the branch lying between the curve of cusps and the limit $\lambda\rightarrow1+\nu$ ($j_\psi=1,j_\phi=0$).

\end{itemize}

Hence we are now in a position to interpret the three parameters of the doubly spinning ring solution, which are as follows:

\begin{itemize}

\item
The parameter $k$ is analogous to $R$ in \cite{ringcoords}, and sets the scale for the solution.  Fixing the mass (as we do throughout) gives $k$ in terms of $\nu$ and $\lambda$.

\item
$\nu$ provides an upper bound on the $S^2$ angular momentum: as $\nu$ increases, the upper bound on the $S^2$ angular momentum also increases.  In the case that $\nu=0$ we find rings with angular momentum only on the $S^1$.

\item
$\lambda$ can be thought of as a shape parameter, with a critical value separating fat and thin ring branches being given by the equation of the curve of cusps:

\be
	\lambda=\frac{1}{4}(-1-\nu+\sqrt{(9+\nu)(1+9\nu)}).
\ee

When $\lambda$ is less than this value, we obtain thinner rings.  Conversely, if $\lambda$ is greater than this value, we retrieve rings on the fat ring branch.

\end{itemize}

\begin{figure}[h]
\begin{center}
\includegraphics[width=15.cm]{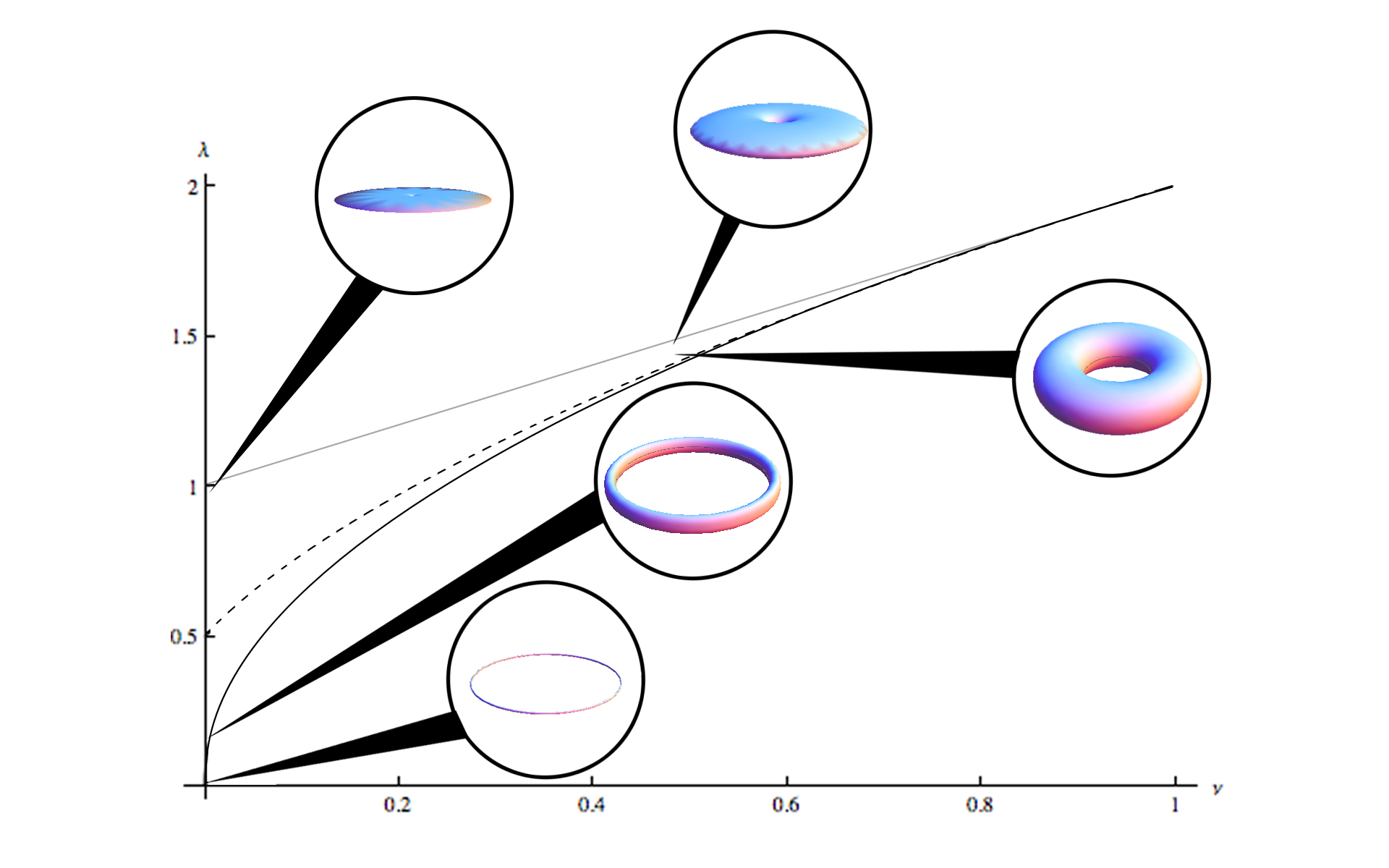}
\end{center}
\caption{\small Parameter space for doubly spinning black rings.  The dashed curve is the curve of cusps, the critical value separating fat and thin ring branches.  The black rings shown are isometric embeddings of the $S^2$ cross section with the size of the $S^1$ estimated by the inner radius of the horizon.}
\label{figure:parameter}
\end{figure}

As the metric naturally lends itself to working in terms of these parameters, the black hole configurations in parameter space are summarized in figure \ref{figure:parameter}.  Two regions of this parameter space diagram are of particular note:

\begin{itemize}

\item
Extremal doubly spinning rings, which have maximum $j_\phi$ for given $j_\psi$, exist on the lowest curve, $\lambda=2\sqrt{\nu}$.

\item
The limit $\lambda=1+\nu$ corresponds to the extremal Myers-Perry black hole, and is given by the upper line.

\end{itemize}
	
\section{Geometry of the ring}
\label{sec:geometry}

A detailed analysis of the shape of singly spinning rings was carried out in \cite{dynamics}, where the horizon deformation from a circular $S^2$ and its relative position from the $\psi$ axis of rotation were made clear.  Our aim is to show how the addition of angular momentum on the $S^2$ affects the horizon geometry and examine what new configurations this allows.

The horizon shape is most easily visualized through the technique of isometric embedding which preserves distances from a curved spacetime to a fiducial Euclidean space (see Appendix A of \cite{dynamics} for a review of the procedure which is adopted here).  In ring coordinates, the black ring metrics readily admit a suitable form for isometric embedding through the substitution $x\rightarrow\cos\theta$.

\begin{figure}[h]
\begin{center}
\includegraphics[width=15cm]{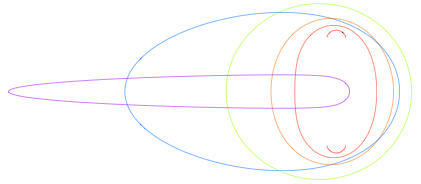}
\begin{eqnarray*}
{\small
~~~~~
\begin{array}{r|r|r|r|r|r|r}
	\lambda & 0.142 & 0.17 & 0.22 & 0.4 & 0.8 & 1.0  \\
 	\hline
 	j_\psi  & 1.13 & 1.07 & 1.00 & 0.92 & 0.95 & 0.999 \\
 	\hline
  	j_\phi & 0.124 & 0.111 & 0.094 & 0.057 & 0.016 & 0.0004 \\
\end{array}
}
\end{eqnarray*}
\end{center}
\caption{\small Cross section of isometric embedding of event horizon for doubly spinning black rings with fixed mass $G_NM=1$ and $\nu=0.005$.  From left to right, the values of $\lambda$ are given by 1.0 (purple), 0.8, 0.4, 0.22, 0.17, 0.142 (red).  The table gives values of the reduced angular momenta for each configuration.}
\label{figure:horizonembed005}
\end{figure}

\begin{figure}[h]
\begin{center}
\includegraphics[width=15cm]{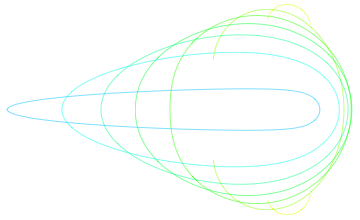}
\begin{eqnarray*}
{\small
~~~~~
\begin{array}{r|r|r|r|r|r|r|r}
 	\lambda & 0.9 & 0.95 & 1.0 & 1.05 & 1.1 & 1.15 & 1.195 \\
  	\hline
  	j_\psi  & 0.78 & 0.81 & 0.85 & 0.89 & 0.93 & 0.96 & 0.996 \\
  	\hline
  	j_\phi & 0.226 & 0.186 & 0.147 & 0.108 & 0.071 & 0.035 & 0.003 \\
\end{array}
}
\end{eqnarray*}
\end{center}
\caption{\small Cross section of isometric embedding of event horizon for doubly spinning black rings with fixed mass $G_NM=1$ and $\nu=0.2$.  From left to right, the values of $\lambda$ are given by 1.195 (light blue), 1.15, 1.1, 1.05, 1.0, 0.95, 0.9 (yellow).}
\label{figure:horizonembed2}
\end{figure}

Embeddings of the event horizon for two different values of $\nu$ are shown in figures \ref{figure:horizonembed005} and \ref{figure:horizonembed2}. In the limit $\nu\rightarrow0$ we retrieve the shapes of the horizon cross-section seen for singly spinning rings: circular for thin rings and very squashed for fat rings.  As the angular momentum in the $\phi$ direction is increased, a thin ring reduces to a rotating black hole cross a line: just as the Kerr horizon becomes oblate, the ring also becomes oblate.  Another feature of the Kerr black hole, the failure of the embedding condition near the axis of rotation, also manifests itself in the doubly spinning black ring.  As one might expect, in the case of thin rings the region which fails to be embeddable is approximately symmetric with respect to both poles, while for fat rings there is asymmetry as there is a clear distinction between the inner and outer rims.
  
\begin{figure}[h]
\begin{center}
\includegraphics[width=4.5cm]{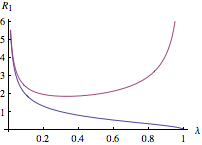}
\hspace{0.5cm}
\raisebox{0cm}{
  \includegraphics[width=4.5cm]{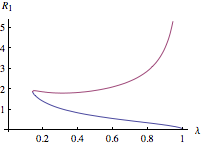}}
\hspace{0.5cm}
\includegraphics[width=4.5cm]{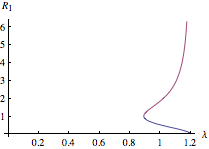}
\end{center}
\caption{\small Inner and outer radii of the event horizon for fixed $\nu$: the plots are given by $\nu=0.00005, 0.005$ and $0.2$ respectively.}
\label{figure:radii}
\end{figure}

While the isometric embedding gives the shape of the horizon, it does not give any information about its location in space.  As such, this information must be obtained by analyzing the metric directly.   The $S^1$ radius at a particular latitude on the horizon is given by
\be
	R_{1}^x=\sqrt{g_{\psi\psi}\left(y=\frac{-\lambda+\sqrt{\lambda^2-4\nu}}{2\nu},x\right)}.
\ee
Ring coordinates are defined such that the inner radius of a ring, $R_{1}^{in}$, lies at $x=1$ and the outer radius, $R_{1}^{out}$, lies at $x=-1$.  Plots of the inner and outer radii are given in figure \ref{figure:radii}.  For all values of $\nu$, $R_{1}^{in}$ decreases as $\lambda$ increases, approaching zero as $\lambda\rightarrow1+\nu$.  The behaviour of the outer radius varies, however, depending on $\nu$.  When $\nu$ is very small (approximately less than 0.009) $R_{1}^{out}$ decreases along the thin ring branch before then increasing, as seen for the singly spinning ring.  For all other values of $\nu$, $R_{1}^{out}$ monotonically increases as $\lambda$ increases.

\begin{figure}[h]
\begin{center}
\includegraphics[width=15cm]{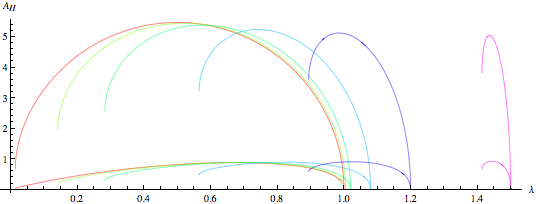}
\end{center}
\caption{\small $S^2$ and total horizon areas for doubly spinning black rings of fixed mass at various fixed $\nu$.  The values of $\nu$ are, from left to right, 0.00005 (red), 0.005, 0.02, 0.08, 0.2, 0.5 (purple).  The upper curves denote the total horizon area while the lower curves represent the $S^2$ horizon area.}
\label{figure:areas}
\end{figure}

However, it is important to remember that in the case of extremal rings, where $R_{1}^{in}$ and $R_{1}^{out}$ coincide, the $S^2$ of the horizon still exists and the ring has non-zero horizon area.  The $S^2$ horizon area is given by
\be
	\mathcal{A}_{S^2}=2\pi\int_{-1}^1 dx\,\sqrt{g_{xx}g_{\phi\phi}}\Big|_{y=\frac{-\lambda+\sqrt{\lambda^2-4\nu}}{2\nu}}.
\ee
A plot of both the $S^2$ and total horizon areas is shown in figure \ref{figure:areas}.  Extremal rings have zero horizon area only in the $j_\phi\rightarrow0$ limit and both their $S^2$ and total horizon areas grow as the $S^2$ angular momentum increases, as expected.  

Therefore, we note the following features of the doubly spinning ring:

\begin{itemize}

\item
Rotation on the $S^2$ causes the horizon to become oblate.  Thin rings have largest $j_\phi$, and hence the effect is most pronounced in these cases.  As such, what were termed `very thin' rings in the case of singly spinning rings are now generally thin only in terms of their $S^1$ radii.

\item
The addition of $S^2$ angular momentum allows very thin rings (in terms of $S^1$ radii) which exist much closer to the axis of rotation.  Physically, this corresponds to the spin-spin interaction of the rotating two-sphere across the centre of the ring: the $S^2$ angular momentum provides an additional effective repulsion to stabilize the ring.

\end{itemize}

\section{Physics of black ring ergosurfaces}
\label{sec:physics}

\subsection{Ergosurface self-merger}
\label{sec:merger}

The addition of angular momentum on the $S^2$ of the ring creates an ergoregion which, under certain conditions, merges with itself across the central axis of rotation.  Hence the topology of the ergosurface changes from $S^1 \times S^2$ to $S^3$.  This process has been illustrated in Weyl coordinates \cite{ergoregion} but we illustrate it here in polar coordinates: these provide an immediately intuitive picture of the occurring physics.  We construct our polar coordinates entirely analogously to \cite{ringcoords} with the exception of scale factor, so that they are defined 
\be
	r_1=k\frac{\sqrt{1-x^2}}{x-y}, \qquad
	r_2=k\frac{\sqrt{y^2-1}}{x-y}.
\ee
The location of the ergosurface is found by setting $g_{tt}=0$.  Thus, in this case, the ergosurfaces are given by the solution of
\be
	{H(y,x)}=1+\lambda^2-\nu^2+2\lambda\nu(1-y^2)x+2y\lambda(1-x^2\nu^2)+y^2x^2\nu(1-\lambda^2-	\nu^2)=0
\ee
which are
\be
	y=\frac{-2\lambda+2x^2\lambda\nu^2\pm\sqrt{(2\lambda-2x^2\lambda\nu^2)^2-4(1+\lambda^2+2x\lambda\nu-\nu^2)(x^2\nu-2x	\lambda\nu-x^2\lambda^2\nu-x^2\nu^3)}}{2(x^2\nu-2x\lambda\nu-x^2\lambda^2\nu-x^2\nu^3)}.
\ee
The solution with positive sign corresponds to the ergosurface that we seek.  The solution with negative sign corresponds to a mathematical inner ergosurface, inside the inner horizon.  This is not to be confused with the inner part of the outer ergosurface, which excludes the origin - the intersection of the $\psi$ and $\phi$ axes - from the ergoregion, as will be seen shortly.

\begin{figure}[h]
\begin{center}
\includegraphics[width=4.5cm]{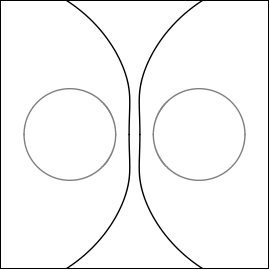}
\hspace{0.5cm}
\raisebox{0cm}{
  \includegraphics[width=4.5cm]{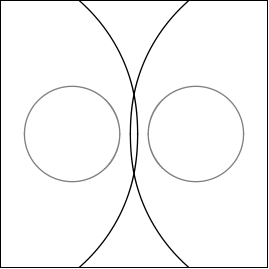}}
\hspace{0.5cm}
\includegraphics[width=4.5cm]{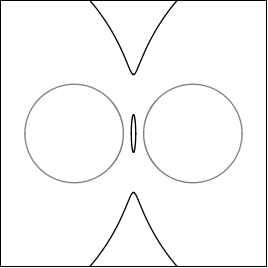}
\end{center}
\caption{\small Cross section of black ring ergosurface merger in polar coordinates.  The grey circles are the event horizon while the black curves are the outer ergosurface. In this figure, $\nu=0.1$ in all three illustrations, while $\lambda=0.85$ ($S^1{\times}S^2$ ergosurface), $\lambda=0.9$ (merger point) and $\lambda=0.95$ ($S^3$ ergosurface) respectively.}
\label{figure:ergosurface}
\end{figure}

When a merger occurs, it must do so on the $\psi$ axis, by symmetry.  In ring coordinates, this axis is given by $y=-1$.  The merger condition can be found by considering the $g_{tt}$ component of the metric along this axis: the point at which this occurs must have $g_{tt}$ instantaneously equal to zero before once again becoming negative.  By considering
\be
	\frac{dH(-1,x)}{dx}=4x\lambda\nu^2+2x\nu(1-\lambda^2-\nu^2)=0,
\ee
we see that the merger point must occur when $x=0$.   As a result, the merger condition is given by
\be
	H(-1,0)=1+\lambda^2-\nu^2-2\lambda=0,
\ee
or more simply,
\be
	\nu+\lambda=1.
\ee
The lower bound on $\lambda$ implies that a transition between ergosurface topologies can take place only when $\nu\le3-2\sqrt{2}$.  An example of such a transition is given in figure \ref{figure:ergosurface}.

In the $S^3$ ergosurface configuration, a region in the centre of the ring is excluded.  This region always includes the origin, due to it being a point of symmetry.  One can think of the rotating black ring as sweeping up the surrounding spacetime as it rotates around the $\psi$ and $\phi$ axes: at the origin all the frame-dragging effects must cancel.  In addition, the area of this region decreases as both $\nu$ and $\lambda$ increase.  This can be interpreted in terms of the $R^{in}_1$ radius.  As $\nu$ increases, even the thinnest rings are less distant from the $\psi$ axis of rotation.  Then, as $\lambda$ increases, the rings increase in fatness, and their $R^{in}_1$ decreases yet further.

The ergosurface merger condition implies that merged ergoregions cannot simply be attributed to thin or fat rings without further consideration.  Clearly the angular momentum on the $S^2$ is the attributable factor for the presence of merged ergoregions as singly spinning ring ergosurfaces are all of topology $S^1{\times}S^2$.  Indeed, this $S^2$ angular momentum can be thought of explicitly by considering the doubly spinning ring as a Kerr black hole at every point on the circle.  In polar Cartesian Kerr-Schild coordinates, the ergosurface of a Kerr black hole bulges away from the horizon when the angular momentum is sufficiently large \cite{kerr}.  With this in mind, the topology of the black ring ergosurface is determined by three factors: 

\begin{itemize}

\item
relative competition between $j_\psi$ and $j_\phi$ angular momenta: if $j_\psi$ is significantly larger than $j_\phi$, the ergosurface tends to have topology $S^1{\times}S^2$,

\item
inner horizon radius, $R^{in}_1$: rings whose inner radii are further from the $\psi$ axis of rotation are less likely to have merged ergoregions,

\item
the thickness of the ring: fat rings have inner radii which tend to the $\psi$ axis of rotation in the limit $\lambda\rightarrow1+\nu$, and hence are more likely to have ergosurfaces of topology $S^3$.

\end{itemize}

\begin{figure}[h]
\begin{center}
\includegraphics[width=4.5cm]{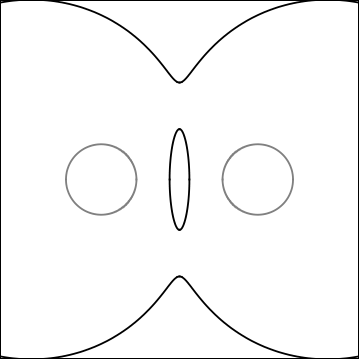}
\hspace{1cm}
\raisebox{0cm}{
  \includegraphics[width=4.5cm]{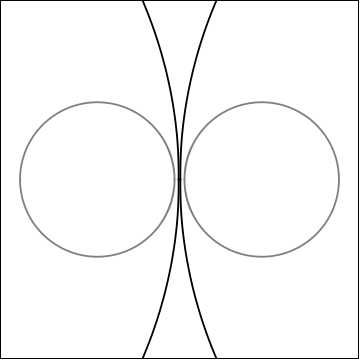}}
\end{center}
\caption{\small Cross section of black rings with atypical ergosurface behaviour.  The illustration on the left depicts a thin ring ($\nu=0.18, \lambda=0.85$) whose ergosurface has merged: the ergosurface cross section is almost circular and the inner radius of the horizon is relatively large.  On the right we see a fat ring ($\nu=0.01, \lambda=0.95$) whose ergosurface has not merged: the ergosurface is extremely large compared to the ring horizon, whose inner radius is very small.}
\label{fig:cases}
\end{figure}

Hence, in general, it is thin black rings that have unmerged ergosurfaces, but there are also a smaller number of thin rings with merged ergosurfaces and fat rings with unmerged ergosurfaces.  Examples\footnote{Curves of constant $y$ are circles in the polar coordinates being used.  As such, the shape of the horizon can be deduced by considering the ergosurface size and the inner radius of the horizon.} of these are shown in figure \ref{fig:cases}.

\subsection{Ergosurface shape and physics}
\label{sec:ergoshape}

Having analyzed the shape of the horizon in detail in section three, we now apply a similar analysis to the ergosurface.  Unfortunately, the metric of the doubly spinning ring does not readily admit the form necessary for isometric embedding of the ergosurface due to it being specified as a complicated function of both $x$ and $y$.  Thus, the analysis presented here is given for the singly spinning ring with angular momentum on the $S^1$ \cite{ringcoords}.  In this case, the ergosurface is located at
\be
	y=-\frac{1}{\lambda}
\ee
or, with substitution of the balance condition,
\be
	y=-\frac{1+\nu^2}{2\nu}.
\ee
The portion of the induced metric required for isometric embedding is given by
\beqa
	ds^2 &=& \frac{2(1-\nu)(1+\nu^2)(1+\frac{2\nu\cos\theta)}{1+\nu^2})\sin^2\theta}{3\pi\nu(\frac{1+\nu^2}{2\nu}+\cos\theta)^2(1+	\nu\cos\theta)(1-\cos^2\theta)}d\theta^2 \nonumber\\
	&& +\frac{2(1+\nu^2)(1-\frac{2\nu}{1+\nu^2})(1+\nu\cos\theta)(1-\cos^2\theta)}{3\pi(1-\nu)\nu(\frac{1+\nu^2}{2\nu}+\cos	\theta)^2}d\phi^2,
\eeqa
where the scale $G_NM=1$ has been fixed, $x\rightarrow\cos\theta$, and the angular coordinate $\phi$ has been rescaled to have periodicity 2$\pi$.  We find that the embedding condition is satisfied everywhere on the ergosurface for all $\nu$.

\begin{figure}[h]
\begin{center}
\includegraphics[width=15cm]{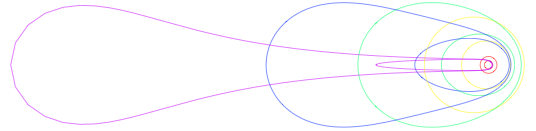}
\end{center}
\caption{\small Cross section of isometric embedding of horizon and ergosurface for singly spinning black rings with fixed mass $G_NM=1$.  The horizon and ergosurface for a particular configuration are given by curves of the same colour.  From left to right, the values of $\nu$ are given by 0.995 (purple), 0.8 (blue), 0.5 (green), 0.2 (yellow) and 0.005 (red).}
\label{fig:horergoembed}
\end{figure}

A cross section of the isometric embedding for different configurations of singly spinning rings is given in figure \ref{fig:horergoembed}.  It is immediately evident that both the shape and relative size of the ergosurface vary greatly dependent on the thickness of the ring: we attribute this to the linear velocity of the event horizon in the plane of the ring.  More precisely, every point on the horizon can be thought to be contributing to the ergosurface being generated.  A point on the horizon which is moving with a greater linear velocity exerts a correspondingly larger frame dragging effect.  

Consider a very thin ring with $\nu\ll1$: its $R_1^{in}$ and $R_1^{out}$ almost coincide, which accounts for the almost symmetric ergosurface cross section.  Its circular shape and small extent are a result of its low angular velocity: since the corresponding linear velocity is low, the small frame dragging affects the surrounding space equally.  Then, as $\nu$ increases, the inner radius of the horizon decreases, while at the same time the angular velocity increases: this accounts for the change in distance between the inner radii of the ergosurface and horizon seen in the figure.  

One curious point of interest is the bulbous nature of the ergosurface further away from the outer horizon, which is particularly pronounced in the case of very fat rings.  It is important to remember that the ergosurface is intimately tied to frame dragging.  With this in mind, this bulbous ergosurface can still be understood in terms of the linear velocity.  At the tip of the outer rim of a fat ring, the linear velocity is very large and is of course tangential to the direction of rotation.  As such, the resulting frame dragging is both severe and immediately following `behind' the point as it rotates.  This effect initially follows radially in the plane of the ring before spreading through space gradually as the gravitational forces decrease.

\section*{Acknowledgements}

We are very grateful to Veronika Hubeny for conversations, comments and suggestions.  This work was supported by a bursary from the Nuffield Foundation, URB/35686.


\end{document}